\documentstyle[sprocl,epsfig]{article}

\def\Journal#1#2#3#4{{#1}{\bf #2} (#4) #3}

\def\NPB{{\em Nucl. Phys.} B}
\def\PLB{{\em Phys. Lett.} B}

\def\ZPC{{\em Z. Phys.} C}

\def\a{\alpha}
\def\b{\beta}

\def\d{\delta}
\def\e{\epsilon}
\def\f{\phi}
\def\g{\gamma}
\def\h{\eta}

\def\j{\psi}

\def\l{\lambda}
\def\m{\mu}
\def\n{\nu}
\def\o{\omega}
\def\p{\pi}

\def\x{\xi}
\def\z{\zeta}

\def\G{\Gamma}

\def\co{{\cal O}}

\def\bo{{\raise.15ex\hbox{\large$\Box$}}}               % D'Alembertian
\def\face{{\raise.2ex\hbox{$\displaystyle \bigodot$}\mskip-2.2mu \llap {$\ddot
        \smile$}}}                                      % happy face
\def\Bar#1{\overline{#1}}                       % big bar
\def\leftrightarrowfill{$\mathsurround=0pt \mathord\leftarrow \mkern-6mu
        \cleaders\hbox{$\mkern-2mu \mathord- \mkern-2mu$}\hfill
        \mkern-6mu \mathord\rightarrow$}       % <--> double differential
\def\dvec#1{\vbox{\ialign{##\crcr
        \leftrightarrowfill\crcr\noalign{\kern-1pt\nointerlineskip}
        $\hfil\displaystyle{#1}\hfil$\crcr}}}           % <--> accent
\def\beq{\begin{equation}}
\def\eeq{\end{equation}}

\def\beqx{\begin{displaymath}}
\def\eeqx{\end{displaymath}}

\def\beql{\begin{eqnarray}}
\def\eeql{\end{eqnarray}}
\def\NO{\nonumber}

\begin{document}

\thispagestyle{empty}
\setcounter{page}{0}
\hfill {\tt hep-ph/9805370}

\hfill {NIKHEF 98-011}
\vskip 42pt
\centerline{\bf SUDAKOV RESUMMATION FOR ELECTROPRODUCTION }
\centerline{OF HEAVY QUARKS}
\vskip 32pt\centerline{S. Moch}
\centerline{\it NIKHEF Theory Group}
\centerline{\it P.O. Box 41882, 1009 DB Amsterdam, The Netherlands}
\vskip 32pt
{\narrower\baselineskip 10pt
\centerline{\bf Abstract}
\noindent
The leading and next-to-leading threshold logarithms of the QCD corrections 
to electroproduction of heavy quarks in single-particle inclusive kinematics 
are resummed to all orders in perturbation theory. 
The resummed cross-section is used to derive the NLO and NNLO results near 
threshold and their numerical impact on the charm structure function 
is studied.
\smallskip}
\vskip 96pt
\centerline{Talk at the }
\centerline{\it `6th International Workshop on Deep Inelastic Scattering and QCD
'}
\centerline{{\it (DIS98)}, Brussels, April 1998}
\medskip
\centerline{\it to be published in the proceedings}
\vskip 43pt
\noindent
{May 1998\hfill}
\newpage

\title{SUDAKOV RESUMMATION FOR ELECTROPRODUCTION OF HEAVY QUARKS}

\author{S. MOCH}

\address{NIKHEF Theory Group\\
P.O. Box 41882, 1009 DB Amsterdam, The Netherlands}

\maketitle\abstracts{
The leading and next-to-leading threshold logarithms of the QCD corrections 
to electroproduction of heavy quarks in single-particle inclusive kinematics 
are resummed to all orders in perturbation theory. 
The resummed cross-section is used to derive the NLO and NNLO results near 
threshold and their numerical impact on the charm structure function 
is studied.}

The deep-inelastic production of heavy quarks is an important reaction, 
which is most prominently used for the extraction of the gluon density 
in the proton. 
Experimental data on charm production in lepton proton collisions 
have been published by e.g. EMC \cite{emc}, H1 and ZEUS \cite{h1zeus}, 
while calculations up to next-to-leading order (NLO) 
provide solid foundations for a theoretical description \cite{lrsvn93}.
As a matter of fact, the charm structure function $F_{2}^{\rm c}$ 
is quite sensitive to the effects of partonic processes close 
to the charm quark pair threshold. 
This is due to the large gluon density $g(x,\m)$ at small 
momentum fractions $x$, which enhances the contribution of 
soft gluon emission near the elastic limit \cite{v96}. 
In this kinematical region, the QCD corrections are dominated by large 
Sudakov double logarithms, which have to be resummed to all orders 
of perturbation theory. 
This task has been performed up to next-to-leading 
logarithmic (NLL) accuracy \cite{lm98}. The technology 
and some results are briefly presentend here.

We study electron proton scattering with the exchange 
of a single virtual photon, $Q^2=-q^2$, 
and a detected heavy quark in the final state, thus 
\beql
\g(q) +\, P(p) &\longrightarrow& {\rm{Q}}(p_1) +\, X\, ,
\label{elecprotscatt}
\eeql
where $X$ denotes any additional hadrons in the final state and $p_1^2 =m^2$. 
The Mandelstam invariants, 
$S^{\prime} = (p + q)^2 + Q^2\,, T_1 = (p - p_1)^2-m^2 $ and  
$U_1 = (q -p_1)^2-m^2$ can be used define $S_4=S^{\prime} + T_1 + U_1$,
which vanishes near the partonic threshold.
The double differential heavy quark structure function $dF_2$ 
associated to the process eq.(\ref{elecprotscatt}) may be written as 
\beql
\frac{d^2 F_{2}(x,S_4,T_1,U_1,Q)}{dT_1 dU_1} &=&\!\!
\sum\limits_{i={\rm q},\Bar{{\rm q}},g}^{}\,\,
\int\limits_{ax}^{1} \frac{dy}{y}\,
\f_{i/P}(y,\m)\, \o_{2 i}\! \left(\frac{x}{y},s_4,t_1,u_1,Q,\m\right)\!,\,\,\,
\label{charmstruc}
\eeql
where $a=1+4 m^2/Q^2$. The $\f_{i/P}$ denote parton distributions in 
the proton at mometum fraction $y$ and $\Bar{{\rm{MS}}}$-mass 
factorization scale $\m$. 
The functions $\o_{2 i}$ describe the underlying hard parton 
scattering processes and depend on associated partonic Mandelstam variables 
$s^{\prime}, t_1,u_1$ and $s_4$, which are derived from eq.(\ref{elecprotscatt}) 
after replacing the proton $P$ by a parton of momentum $k=(x/y) p$. 
At $n$-th order in perturbation theory, the gluonic hard part $\o_{2g}$ 
in eq.(\ref{charmstruc}) typically depends on singular distributions 
$\a_s^n [\ln^{2n-1}(s_4/m^2)/s_4]_+$, that have to be resummed. 
Light initial state quarks will be neglected. 

In order to perform this resummation of threshold logarithms, 
the phase space near the elastic edge is decomposed into various regions, 
each depending on an infrared safe kinematical weight $w \simeq 0$. 
For $d F_2$ in eq.(\ref{charmstruc}), this decomposition implies 
$S_4/m^2 = w_1 + w_s$, with $w_1=(1 - x/y) (-u_1)/m^2$ and 
$w_s = s_4/m^2$. Then, general arguments \cite{cls96}
allow for a refactorization of $d F_2$, 
\beql
\frac{d^2 F_{2}(x,S_4,T_1,U_1,Q)}{dT_1 dU_1} &=& 
H_{2 g}(S,T_1,U_1)\, \int dw_1\, dw_s\, 
\d\,\left(\frac{S_4}{m^2} - w_1 - w_s \right) \NO\\
& &\hspace*{10mm} \times\, \j_{g/g}(w_1,p\cdot\z)\, S(w_s,\z,\m)\, ,
\label{threshfac}
\eeql
separating the partonic degrees of freedom for $S_4/m^2 \to 0$ 
into different functions of individual weight $w$. 
The hard function $H_{2g}$ summarizes off-shell short-distances corrections, 
while $\j_{g/g}$ accounts for effects of collinear gluons 
and the soft function $S$ for soft, long-wavelength gluons. 
Replacing the proton in eq.(\ref{charmstruc}) by a gluon and taking 
Laplace moments, 
$\tilde{f}(N) = \int_0^{\infty}\! dw\, {\rm{exp}}[-N w] f(w)$, gives
\beql
{\tilde{\o}}_{2 g}(N,t_1,u_1)
&=& H_{2 g}(S,T_1,U_1) 
\left[\frac{\tilde{\j}_{g/g}(N,p\cdot\z)}{\tilde{\f}_{g/g}(N,\m)}\right]
{\tilde{S}}(N,\z,\m)\, + \co(1/N)\,.
\label{omegamom}
\eeql
where $\f_{g/g}$ is the usual $\Bar{\rm MS}$-distribution 
from mass factorization. In moment space, the Sudakov logarithms appear 
as factors $\a_s^n \ln^{2n-i}\!\!N$, with $i=0,1$ for NLL accuracy, and  
the $N$-dependence in eq.(\ref{omegamom}) exponentiates for each function 
individually. All leading logarithms (LL) are exclusively contained 
in ${\tilde{\j}}_{g/g}$, which is a gluon distribution at fixed energy defined 
as an operator matrix element \cite{s87}. 
It depends on a time-like vector $\z$, $\z^2=1$, that fixes the kinematics 
by projecting on the proper energy fraction $p\cdot \z$ of the proton.  
At order $\a_s$ in $D=4-2\e$ dimensions and axial gauge $n \cdot A = 0$, 
$n=\z$, $\j_{g/g}$ is given by 
\beql
\j_{g/g}(w,p\cdot \z) = \frac{\a_s}{\p} C_A \!\left\{ \! \frac{-1}{\e} 
\bigl[\frac{1}{w} \bigr]_+ + \bigl[\frac{2 \ln w}{w} \bigr]_+
+ \bigl[\frac{1}{w} \bigr]_+ \! \left( \! \ln \frac{4 (p \cdot \z)^2}{\m^2}
- 1  \! \right) \! \right\} ,\,\,\,\,
\eeql
where the remaining collinear poles are cancelled by $\f_{g/g}$. 
The LL logarithms in ${\tilde{\j}}_{g/g}$ are resummed 
in analogy to the Drell-Yan process \cite{s87}, while all scale dependence 
of ${\tilde{\j}}_{g/g}$  and ${\tilde{\f}}_{g/g}$ is governed by 
renormalization group equations (RGE) 
with anomalous dimensions $\g_\j$ and $\g_{g/g}$ \cite{kos98}.

The soft function $S$ requires renormalization, 
since it is defined as a composite operator, that connects Wilson 
lines in the direction of the scattering partons \cite{cls96,ks97}. 
Its RGE, $\m (d/d\m)  \ln \tilde{S}(N) = - 2\, {\rm{Re}} \G_S$,
resums all logarithms in $\tilde{S}$ and its gauge dependence cancels 
precisely the corresponding terms in $\j_{g/g}$. 
The soft anomalous dimension $\G_S$ is a one-dimensional 
matrix in colour space and to order $\a_s$ calculated to be
\beql
\G_S(\a_s)&=&\frac{\a_s}{\p}
\Biggl\{ \!\left(\frac{C_A}{2} - C_F\right)\! ( L_\b + 1 ) - 
\frac{C_A}{2} \left( \ln (p \cdot \z)^2 + 
\ln\frac{4\, m^2}{t_1\, u_1} \right)\! \Biggr\}\, ,
\label{softadim-res}
\eeql
with $\b = \sqrt{1- 4\, m^2/s}$ and 
$L_\b=(1-2\,m^2/ s)\{ \ln (1-\b)/(1+\b) + {\rm{i}}\p \}/\b$.

The final result for the hard scattering function  
$\tilde{\o}_{2g}$ in moment space resums all large logarithms 
in single-particle inclusive kinematics up to NLL accuracy \cite{los98}. 
Combining the resummed $\tilde{\j}_{g/g}$ with the integrated 
RGE for $\tilde{S}$, 
we obtain for $\tilde{\o}_{2g}$
\beql
{\lefteqn{
{\tilde{\o}}_{2 g}(N,t_1,u_1) = H_{2 g}(S,T_1,U_1)\,  }}\\
& &\times \,  
{\rm exp}\Bigl\{ \int\limits_0^{\infty}\! \frac{d w}{w}  
\! \left(1 - {\rm{e}}^{-Nw} \right)\! 
\Bigl[\, \int\limits_{w^2}^1 \frac{d \l}{\l} 
A_{(g)}(\a_s(\l m^2)) + \frac{1}{2} \n_{(g)}(\a_s(w^2 m^2)) \Bigr] 
\Bigr\} \NO\\
& &\times \,  
{\rm exp}\Bigl\{2\!\!\!\! \int\limits_\m^{p \cdot \z/N}\!\!  \frac{d\l}{\l}
{\rm Re}\G_S(\a_s(\l^2)) \Bigr\} \,
{\rm exp}\Bigl\{- 2\! \int\limits_\m^{p \cdot \z}\!\!  \frac{d\l}{\l}
\left(\g_g(\a_s(\l^2)) - \g_{g/g}(\a_s(\l^2)) \right) \Bigr\} \,.\NO
\label{omegaresum}
\eeql
The first exponent gives the leading $N$-dependence of the ratio 
${\tilde{\j}}_{g/g}/ {\tilde{\f}}_{g/g}$ 
with $\n_{(g)}(\a_s) = 2 C_A \a_s/\p$, 
$A_{(g)}(\a_s) = C_A (\a_s/\p) + (C_A K/2) (\a_s/\p)^2$  and 
$K=C_A(67/18-\p^2/6)-5/9n_f$, the latter ones being the well-known \cite{kt82}.

\begin{figure}[t]
\begin{center}
\epsfig{file=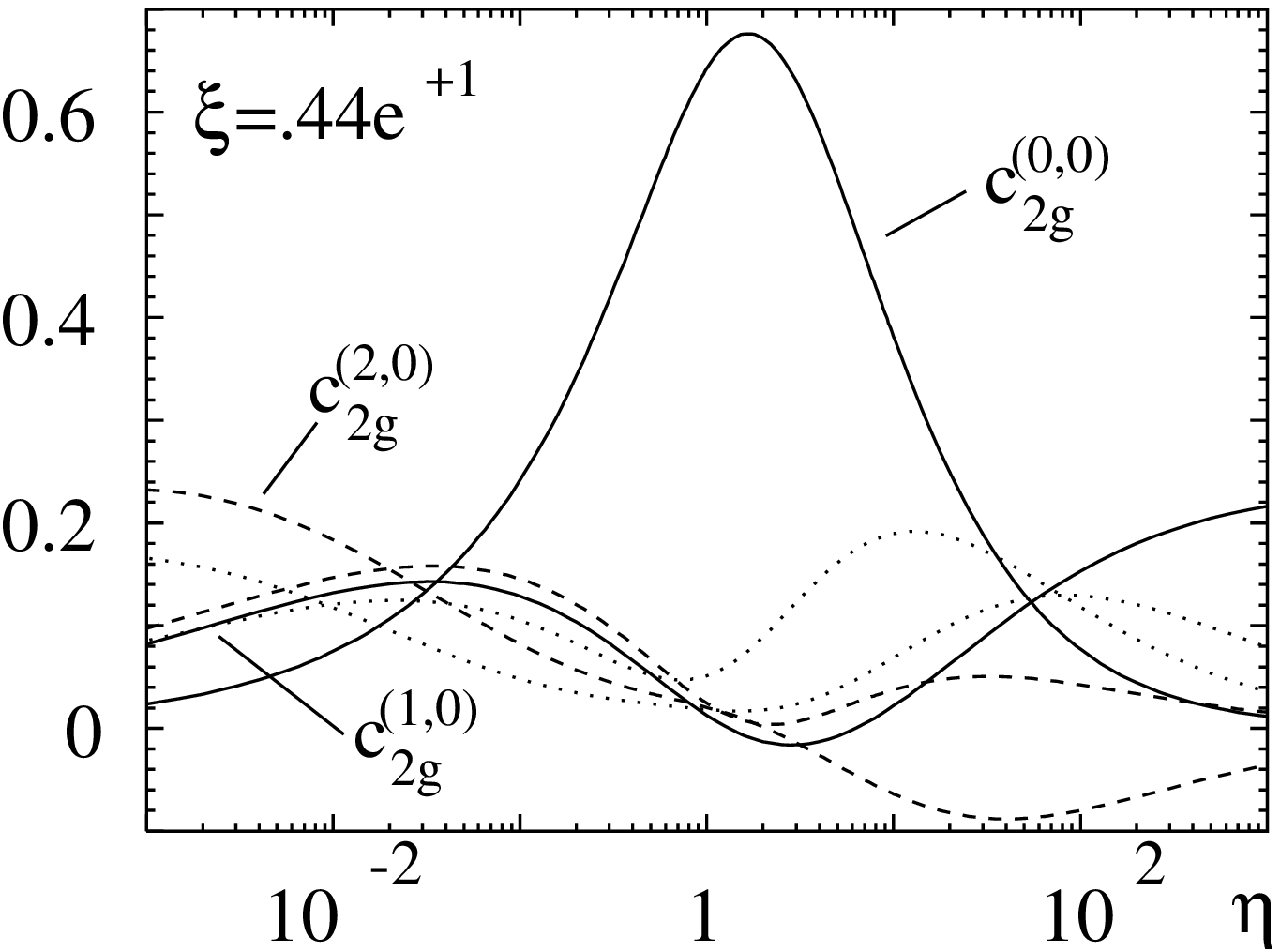,width=5.5cm}
\epsfig{file=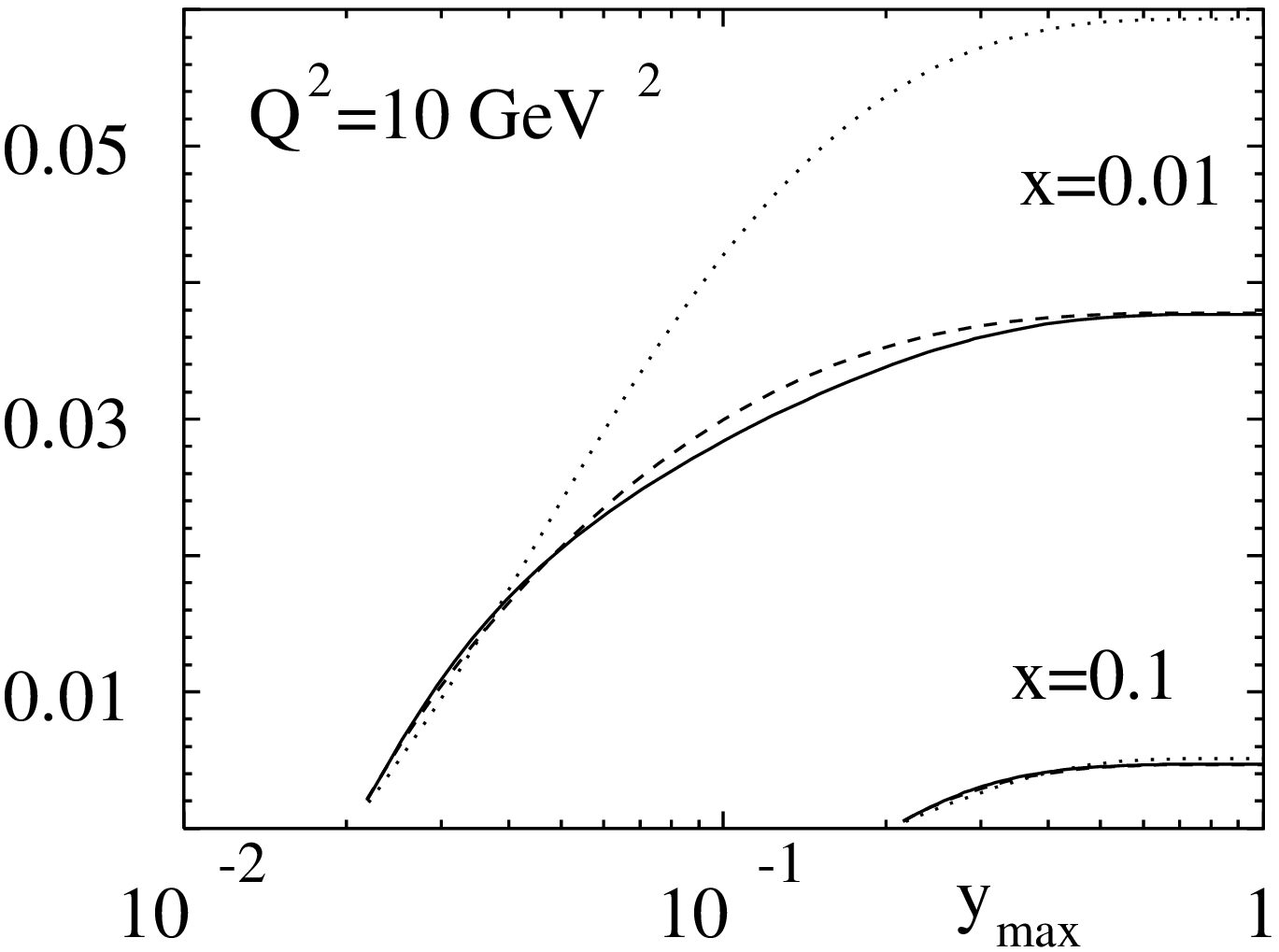,width=5.5cm}
\caption{{\small{Left: 
The $\h$-dependence of the gluon coefficient functions; 
exact results for $c^{(0,0)}_{2g}$ and $c^{(1,0)}_{2g}$ (solid lines); 
approximate LL and NLL results for $c^{(1,0)}_{2g}(c^{(2,0)}_{2g})$ 
(dotted and dashed lines). 
Right: Threshold dependence of the charm structure function $F^{\rm c}_{2}$ 
at NLO with the CTEQ4M gluon distribution and 
$\m=m=1.6 {\rm GeV}$; exact results (solid lines); approximate LL and NLL 
results (dotted and dashed lines).
\label{fig1}}}}
\end{center}
\end{figure}

Using the resummed result for $\tilde{\o}_{2g}$ in eq.(\ref{omegaresum}) 
as a generating functional, we reexpand $\tilde{\o}_{2g}$ to NLO and NNLO.
Inverting the Laplace transform and 
integrating over the phase space in eq.(\ref{charmstruc}) leaves us 
for $F_2$ with 
\beql
F_{2}(x,Q) \!&\simeq&\!
\frac{\a_s(\m^2)\, e_{\rm Q}^2 Q^2}{4 \p^2 m^2}\! 
\int\limits_{ax}^{1}\, dy g(y,\m) 
\sum\limits_{k=0}^{\infty} (4 \p \a_s(\m^2))^k 
\sum\limits_{l=0}^{k} 
c^{(k,l)}_{2 g}(\h,\x) \ln^l\frac{\m^2}{m^2} , \,\,\,\,\,
\label{charmstrucintegrated}
\eeql
where $c^{(k,l)}_{2g}$ are the standard gluon coefficient 
functions \cite{lrsvn93}, $\h = (\x/4) (y/x-1)-1$ and $\x = Q^2/m^2$.
In fig.\ref{fig1} on the left, we plot $c^{(k,0)}_{2g}$ 
as a function of $\h$ for $\x=4.4$ and compare the 
LL and NLL approximations up to NNLO.
On the right, we plot the charm structure function 
$F^{\rm c}_{2}$ for $x=0.1(0.01)$, 
$Q^2=10 \rm{GeV}^2$ as a function of a $y_{max}$-cut on the integral 
in eq.(\ref{charmstrucintegrated}) to estimate threshold sensitivity. 

In general, at  order $\a_s$ the NLL logarithms approximate 
$F^{\rm c}_{2}$ very well, for moderate values of $Q^2$,
better than the LL logarithms, 
while the result for the gluon coefficient function 
$c^{(2,0)}_{2g}$ represents the best present estimate 
of the NNLO corrections. 
An extensive derivation of the result eq.(\ref{omegaresum}) 
for $\tilde{\o}_{2g}$ and a detailed study of its phenomenological 
consequences will be given elsewhere \cite{lm98}.

%%%%%%%%%%%%%%%%%%%%%%%%%%%%%FIGURE  %%%%%%%%%%%%%%%%%%%%%%%%%%%%%%%%%%%
%%%%%%%%%%%%%%%%%%%%%%%%%%%%%%%%%%%%%%%%%%%%%%%%%%%%%%%%%%%%%%%%%%%%%%%

\section*{Acknowledgments}
I would like to thank Eric Laenen for a fruitful collaboration on this project.

\end{document}